\documentstyle[times]{article}

\begin{document}

\begin{center}
Division Algebras; Spinors; Idempotents; The Algebraic Structure of Reality \\
\vspace{.25in}
Geoffrey Dixon \\
gdixon@7stones.com  \\
\vspace{.25in}
$\prec$ Last modified: 2010.01.06 $\succ$

\vspace{.25in}
\begin{quotation}
A carefully constructed explanation of my connection of the real normed division algebras to the particles, charges and fields of the Standard Model of quarks and leptons provided to an interested group of attendees of the 2nd Mile High Conference on Nonassociative Mathematics in Denver in 2009.06.

\end{quotation}
\end{center}
\vspace{.25in}

\textbf{Spinors: a la moi} \\

In this article the spinor is at the root of everything.  My notion of what a spinor is derives from Ian Porteous's book \textit{Topological Geometry} which I was directed to some 30 years ago.  Ian presented a table of representations of universal Clifford algebras of p-time:q-space dimensional spacetimes in terms of the first three real normed division algebras: \textbf{R}, \textbf{C}, and \textbf{H} (the remaining division algebra, the octonions, \textbf{O}, will enter shortly).  By the way, I will also employ Ian's notation \textbf{K}(n) to be the algebra of n$\times$n matrices over an algebra \textbf{K} (by a purer breed of mathematicians this is denoted ${\cal{M}}_{n}(\bf{K})$, I believe, which tells you how comfortable I am with that notation).

Let ${\cal{CL}}(p,q)$ be the Clifford algebra of p,q-spacetime (actually, timespace), then any of Ian's representations can be derived from the sequences for p = 0,1,2,3,..., and q = 0 (line 1), and the sequence for p = 0 and q = 0,1,2,3,... (line 2),
$$
\begin{array}{ccccccccccc}
\bf{R} & \bf{R}^{2} & \bf{R}(2) & \bf{C}(2) & \bf{H}(2) & \bf{H}^{2}(2) & \bf{H}(4) & \bf{C}(8) & \bf{R}(16) & \bf{R}^{2}(16) & ... \\

\bf{R} & \bf{C} & \bf{H} & \bf{H}^{2} & \bf{H}(2) & \bf{C}(4) & \bf{R}(8) & \bf{R}^{2}(8) & \bf{R}(16) & \bf{C}(16) & ... \\
\end{array}
$$
and the rule 
$$
{\cal{CL}}(p+1,q+1) = {\cal{CL}}(p,q) \otimes \bf{R}(2).
$$
(There is also a periodicity (Bott) of order 8 indicated in the rows above.  Also, \textbf{R}(2), being isomorphic to ${\cal{CL}}(1,1)$, provides a great way of adding time and a transverse space dimension to a pure (longitudinal) space Clifford algebra.)

Some things to point out in particular: \\
$\bullet$ In each case we can find a set of p+q anticommuting elements of ${\cal{CL}}(p,q)$ (the 1-vectors) the squares of which are $\pm I$, with $I$ the identity of ${\cal{CL}}(p,q)$ (p +, and q -); \\
$\bullet$ The product of these p+q anticommuting 1-vectors is not a real multiple of the identity (this is the 'universal' part of 'universal Clifford algebra'); \\
$\bullet$ The spinor space of ${\cal{CL}}(p,q)$ is the obvious set of n$\times$1 column matrices over \textbf{R}, \textbf{C}, or \textbf{H}, on which our respective representations of ${\cal{CL}}(p,q)$ would most naturally act via left matrix multiplication; \\
$\bullet$ If the underlying division algebra is \textbf{H}, then multiplication on the spinor space by elements of \textbf{H} from the right is an algebra of actions on that space that is not accounted for by the elements of ${\cal{CL}}(p,q)$, and so it is \textit{internal} with respect to those \textit{external} Clifford algebra actions (in the sense that isospin $SU(2)$ is an \textit{internal} symmetry, and in what follows it is from this right action by \textbf{H} that isospin $SU(2)$ arises in the form of the subset of unit elements, which is multiplicatively closed); \\
$\bullet$ And finally, the set of elements generated by taking the commutators of pairs of 1-vectors is the set of 2-vectors, and with respect to the commutator product this set is isomorphic to the Lie algebra $so(p,q) \simeq spin(p,q)$.

One last thing to note: each of the n components of these n$\times$1 spinor columns is an element of \textbf{R}, \textbf{C}, or \textbf{H}, a division algebra.  Conventionally ${\cal{CL}}(3,0)$ is represented by the algebra $\bf{C}(2)$, but 
$$
\bf{C}(2) \simeq \bf{P} \equiv \bf{C}\otimes\bf{H}, 
$$
the complexified quaternions.  These two versions of ${\cal{CL}}(3,0)$ have different spinor spaces.  In the former case the spinor space is the 4-dimensional (over \textbf{R}) and consisting of 2$\times$1 complex matrices; and in the latter case the spinor space is the 8-dimensional 1-component set $\bf{C}\otimes\bf{H}$ itself.  In this latter case the algebra of actions of \textbf{H} multiplication on a spinor from the right is again not accounted for and commutes with the Clifford algebra actions.  In addition, in this case the single spinor component is not an element of a division algebra, but of $\bf{C}\otimes\bf{H}$, which has a nontrivial decomposition of its identity into a pair of mutually orthogonal idempotents that sum to 1.  More on this kind of thing very soon: it is the key to almost everything. \\

\textbf{Octonions as Spinors} \\

\textbf{O} is an 8-dimensional real algebra, and despite its nonassociativity it can be incorporated into this Clifford algebra and spinor scheme.  First some notation:
$$
\forall \; x,w \in {\bf{O}}, \;\;{\mbox{L}}_{x}[w] = xw, \;\; {\mbox{R}}_{x}[w] = wx. 
$$
However, although this notation is somewhat conventional, in all my previous work I've used the following notation (thereby avoiding subscripts on subscripts):
$$
x_{L} \equiv {\mbox{L}}_{x}, \;\; x_{R} \equiv {\mbox{R}}_{x}.
$$
In particular, I use a basis $e_{a}, \; a = 0,1,...,7$, for \textbf{O}, with $e_{0} = 1$ the identity, and 
$$
\{e_{0}, e_{1+k}, e_{2+k}, e_{4+k}\}
$$ 
is a basis for a quaternionic subalgebra for all integers $k$, where the index summation is modulo 7, from 1 to 7.  Define 
$$
e_{Lab...c} \equiv {\mbox{L}}_{e_{a}}{\mbox{L}}_{e_{b}}...{\mbox{L}}_{e_{c}}, \;\; e_{Rab...c} \equiv {\mbox{R}}_{e_{c}}...{\mbox{R}}_{e_{b}}{\mbox{R}}_{e_{a}},
$$
(note reversal of indices in second case), and let ${\bf{O}}_{L}$ and ${\bf{O}}_{R}$ be the algebras spanned by these respective sets of left and right actions.  Finally, and most importantly, 
$$
{\bf{O}}_{L} = {\bf{O}}_{R} \simeq {\bf{R}}(8), 
$$
so these are each the full algebras of endomorphisms on 8-dimensional \textbf{O}.  Any element of ${\bf{O}}_{L}$ can be expressed as a linear combination of elements of ${\bf{O}}_{R}$, and visa versa.  More on this later.

So, 
$$
{\bf{O}}_{L} \simeq {\bf{R}}(8) \simeq {\cal{CL}}(0,6), 
$$
and the spinor space of this representation of ${\cal{CL}}(0,6)$ as ${\bf{O}}_{L}$ is just \textbf{O} itself, which, unlike ${\bf{R}}^{8}$, has a natural multiplicative structure.  The spinor space itself is a division algebra.  (Note: ${\bf{O}}_{L}$ is trivially associative.)  We can represent a basis for the Clifford algebra 1-vectors in this case as 
$$
e_{Lp}, \; p=1,2,3,4,5,6.
$$
The set of 2-vectors is then spanned by 
$$
e_{Lpq}, \; p,q\in\{1,2,3,4,5,6\}, \; p\ne q,
$$
and given the commutator product this is the Lie algebra $so(6)$.  The 6-vector is 
$$
\prod_{p=1}^{p=6}\mbox{L}_{e_{p}} = \mbox{L}_{e_{7}} = \prod_{p=1}^{p=6}e_{Lp} = e_{L7}.
$$

John Huerta (John Baez's student) very kindly pointed out a few egregious errors in the original  version of this article.  He asked for an explicit presentation of my multiplication table.  This is how I write it for myself whenever I need a quick reference:
$$
\begin{array}{c} 
124 \\
235 \\
346 \\
457 \\
561 \\
672 \\
713
\end{array}
$$
These are the 7 sets of "quaternionic" index triples.  For example, from this I deduce that $e_{6}e_{1} = -e_{1}e_{6} = e_{5}$.  In general, if $a$ and $b$ are distinct indices from 1 to 7, then $e_{a}e_{b}$ will be equal to $\pm e_{c}$ for some other index $c$, the sign positive if $b-a$ is a power of 2, and negative otherwise ($b-a$ taken modulo 7, from 1 to 7, so 2-5 = 4, and therefore $e_{5}e_{2}$ is positive ($e_{3}$)). \\

\textbf{Complexified Octonions as Spinors} \\

Define 
$$
\bf{S} = {\bf{C}}\otimes{\bf{O}}.
$$
Since ${\bf{O}}_{L} = {\bf{O}}_{R}$, and trivially ${\bf{C}}_{L} = {\bf{C}}_{R}$, the algebra of left or right  actions of \textbf{S} on itself is 
$$
{\bf{S}}_{L} = {\bf{S}}_{R} = \bf{C}\otimes{\bf{O}}_{L} \simeq \bf{C}\otimes \bf{R}(8) = \bf{C}(8) \simeq {\cal{CL}}(7,0).
$$
Quickly then, so we can get to the res, ${\bf{S}}_{L}$ viewed as the Clifford algebra ${\cal{CL}}(7,0)$ has the following natural identifications: 
$$
\begin{array}{ll}
\mbox{1-vectors: } & ie_{La}, \; a \in \{1,...,7\}, \\
\mbox{2-vectors: }\; (so(7)) & e_{Lab}, \; a, b \in \{1,...,7\}, a\ne b,\\
... & \\
\mbox{7-vector }& \prod_{a=1}^{a=7}i\mbox{L}_{e_{a}} = -ie_{L1234567} = i. \\
\end{array}
$$

The spinor space in this case is \textbf{S} itself, and as was true of the previous case, this spinor space has an algebraic structure of its own.  However, in the previous case the spinor space, \textbf{O}, was a division algebra;  \textbf{S} is not, and it admits a nontrivial resolution of its identity into a pair of orthogonal projectors (idempotents, as long as everything is alternative).  These are 
$$
\rho_{\pm} = \frac{1}{2}(1 \pm ie_{7}) 
$$
(this selection is clearly not unique, but dates back almost 40 years in the literature, and is, given my choice of octonion multiplication table, rather natural).  

The presence of these projectors means there is a natural (ok, I'm over-using that word) decomposition of the spinor space \textbf{S} into 4 mutually orthogonal subspaces:
$$
\begin{array}{l}
{\bf{S}}_{++} = \rho_{+}{\bf{S}}\rho_{+} = \rho_{L+}\rho_{R+}[{\bf{S}}], \;\; \mbox{1-d over } \; \bf{C}\\
{\bf{S}}_{+-} = \rho_{+}{\bf{S}}\rho_{-} = \rho_{L+}\rho_{R-}[{\bf{S}}], \;\; \mbox{3-d over } \; \bf{C}\\
{\bf{S}}_{-+} = \rho_{-}{\bf{S}}\rho_{+} = \rho_{L-}\rho_{R+}[{\bf{S}}], \;\; \mbox{3-d over } \; \bf{C}\\
{\bf{S}}_{--} = \rho_{-}{\bf{S}}\rho_{-} = \rho_{L-}\rho_{R-}[{\bf{S}}], \;\; \mbox{1-d over } \; \bf{C}\\
\end{array}
$$
where $\rho_{L\pm} = \frac{1}{2}(1 \pm ie_{L7})$, and $\rho_{R\pm} = \frac{1}{2}(1 \pm ie_{R7})$, which provides an expression of this decomposition in terms of projectors in ${\cal{CL}}(7,0)$.  

These four reductions of \textbf{S} into four orthogonal subspaces have corresponding reductions of ${\cal{CL}}(7,0)$ into subalgebras that map the four subspaces to themselves.  These are:
$$
\begin{array}{l}
{\cal{CL}}(7,0) \longrightarrow \rho_{L+}\rho_{R+}{\cal{CL}}(7,0)\rho_{L+}\rho_{R+} = {\cal{CL}}_{\rho}(7,0)\rho_{L+}\rho_{R+}, \\
{\cal{CL}}(7,0) \longrightarrow \rho_{L+}\rho_{R-}{\cal{CL}}(7,0)\rho_{L+}\rho_{R-} = {\cal{CL}}_{\rho}(7,0)\rho_{L+}\rho_{R-}, \\
{\cal{CL}}(7,0) \longrightarrow \rho_{L-}\rho_{R+}{\cal{CL}}(7,0)\rho_{L-}\rho_{R+} = {\cal{CL}}_{\rho}(7,0)\rho_{L-}\rho_{R+}, \\
{\cal{CL}}(7,0) \longrightarrow \rho_{L-}\rho_{R-}{\cal{CL}}(7,0)\rho_{L-}\rho_{R-} = {\cal{CL}}_{\rho}(7,0)\rho_{L-}\rho_{R-}, \\
\end{array}
$$
where the subalgebra ${\cal{CL}}_{\rho}(7,0)$ is the same for all four reductions, so we will just look at the (++)-reduction.  (Why are the  ${\cal{CL}}_{\rho}(7,0)$ the same?  In each case the reduction occurs when one of the $\rho$'s goes through the Clifford algebra.  If $e_{L7}$ ($e_{R7}$) anticommutes with a piece of ${\cal{CL}}(7,0)$, then $\rho_{L\pm}$  ($\rho_{R\pm}$) will change to $\rho_{L\mp}$  ($\rho_{R\mp}$) when drawn from one side of that piece to the other, and when it gets there it will encounter $\rho_{L\pm}$  ($\rho_{R\pm}$), and the resulting product is zero, so that piece will be "reduced" out.  So the sign in $\rho_{L\pm}$  ($\rho_{R\pm}$) is immaterial.)  Note first that 
$$
\begin{array}{lll}
\rho_{L\pm}\rho_{L\pm} = \rho_{L\pm} & \Rightarrow & \rho_{L\pm}e_{La}\rho_{L\pm} = e_{La}\rho_{L\pm}, a = 0,7, \\
\rho_{L\pm}\rho_{L\mp} = 0 & \Rightarrow & \rho_{L\pm}e_{La}\rho_{L\pm} = e_{La}\rho_{L\mp}\rho_{L\pm} = 0, a = 1,...,6, \\
\rho_{R\pm}\rho_{R\pm} = \rho_{R\pm} & \Rightarrow & \rho_{R\pm}e_{Ra}\rho_{R\pm} = e_{Ra}\rho_{R\pm}, a = 0,7, \\
\rho_{R\pm}\rho_{R\mp} = 0 & \Rightarrow & \rho_{R\pm}e_{Ra}\rho_{R\pm} = e_{Ra}\rho_{R\mp}\rho_{R\pm} = 0, a = 1,...,6. \\
\end{array}
$$
Therefore, of the seven 1-vectors of ${\cal{CL}}(7,0)$, the only one that survives the reduction to ${\cal{CL}}_{\rho}(7,0)$ is $ie_{L7}$. (Oh, and by the way, $\rho_{L\pm}$ commutes with $\rho_{R\pm}$.)
 
However, what we're really interested in is what happens to the 2-vectors, viewed as a representation of the Lie algebra $so(7)$.  There are 21 elements, $e_{Lab}, \; a,b \in \{1,2,3,4,5,6,7\}$ distinct, but we can divide these into two types: those for which one of the indices is 7; and those for which neither index is 7.  In what follows it will be understood that any index $p,q,r,s \in \{1,2,3,4,5,6\}$.  Ok, so 
$$
\begin{array}{lll}
\rho_{L\pm}e_{Lp7}\rho_{L\pm} = \rho_{L\pm}e_{Lp}e_{L7}\rho_{L\pm} = e_{Lp}\rho_{L\mp}e_{L7}\rho_{L\pm} = e_{Lp}e_{L7}\rho_{L\mp}\rho_{L\pm} =  0, \\
\rho_{L\pm}e_{Lpq}\rho_{L\pm} = \rho_{L\pm}e_{Lp}e_{Lq}\rho_{L\pm} = e_{Lp}\rho_{L\mp}e_{Lq}\rho_{L\pm} = e_{Lp}e_{Lq}\rho_{L\pm}\rho_{L\pm} =  e_{Lpq}\rho_{L\pm}, \\
\end{array}
$$
where the subalgebra of $so(7)$ generated by elements $e_{Lpq}, \; p,q \in \{1,...,6\}$ distinct, is $so(6)$.  But we're not quite done, since we still have to finish the reduction by looking at $\rho_{R\pm}e_{Lpq}\rho_{L\pm}\rho_{R\pm}$.  Since the $\rho_{L\pm}$ is irrelevant, we'll leave it out for now and just look at the elements $\rho_{R\pm}e_{Lpq}\rho_{R\pm}$.  

Once again we're going to divide these 15 index combinations into 2 sets: those for which $e_{p}e_{q} = \pm e_{7}$; and those for which $e_{p}e_{q} = \pm e_{r}, \; r \ne 7$.  We'll consider the latter case first, and there are 12 distinct elements (to within a sign).  We need only look at one, which will be $e_{L12}$ (note: $e_{1}e_{2} = e_{4}$).  Recall, \textbf{O}$_{L}$ = \textbf{O}$_{R}$, so we can express any element of \textbf{O}$_{L}$ as a linear combination of elements of \textbf{O}$_{R}$.  In particular, given the multiplication table employed here (and you'll have to take my word for this, or consult my book or previous papers):
$$
e_{L12} = \frac{1}{2}(e_{R4} - e_{R12} + e_{R63} + e_{R57}).
$$
Therefore, 
$$
\begin{array}{ll}
\rho_{R\pm}e_{L12}\rho_{R\pm} &= \frac{1}{2}\rho_{R\pm}(e_{R4} - e_{R12} + e_{R63} + e_{R57})\rho_{R\pm} \\
 & = \frac{1}{2}( 
  \rho_{R\pm}\rho_{R\mp}e_{R4} 
  - \rho_{R\pm}\rho_{R\pm}e_{R12} 
  + \rho_{R\pm}\rho_{R\pm}e_{R63} 
  + \rho_{R\pm}\rho_{R\mp}e_{R57}) \\
 & = \frac{1}{2}\rho_{R\pm}(-e_{R12} + e_{R63}) \\
 & = \frac{1}{2}(e_{L12} - e_{L63})\rho_{R\pm}. \\
\end{array}
$$
Note: $\rho_{R\pm}$ commutes with $(e_{L12} - e_{L63}) = (-e_{R12} + e_{R63})$.  The other eleven $e_{Lpq}$, such that  $e_{p}e_{q} = \pm e_{r}, \; r \ne 7$, reduce in like fashion when surrounded with $\rho_{R\pm}$, and these 12 elements are not linearly independent.  For example,
$$
\rho_{R\pm}e_{L63}\rho_{R\pm} =  -\frac{1}{2}(e_{L12} - e_{L63})\rho_{R\pm}.
$$
So, in fact there are only 6 independent elements surviving the reduction ($\rho_{R\pm}...\rho_{R\pm}$) of these 12 elements.  These are 6 of the 8 elements of the $su(3)$ Lie algebra that generate an $SU(3)$ subgroup of the Lie group $G_{2}$, the automorphism group of \textbf{O}, that leave the unit $e_{7} \in \bf{O}$ invariant.

The final 3 elements of $so(6)$ we need look at are the $e_{Lpq}$ for which $e_{p}e_{q} = \pm e_{7}$.  These are 
$$
\begin{array}{l}
e_{L13} = \frac{1}{2}(e_{R7} - e_{R13} + e_{R26} + e_{R45}), \\
e_{L26} = \frac{1}{2}(e_{R7} + e_{R13} - e_{R26} + e_{R45}), \\
e_{L45} = \frac{1}{2}(e_{R7} + e_{R13} + e_{R26} - e_{R45}), \\
\end{array}
$$
and (I hope this is obvious) $\rho_{R\pm}$ commutes with every term on the right hand side of these equations, so there is no further reduction achieved at this point.  However, we can take linear combinations of these 3 elements to make it clearer what the overall structure of $\rho_{R\pm}so(6)\rho_{R\pm}$ actually is.  In particular, there are 2 linearly independent elements we get by taking the differences of these 3 in pairs.  Together with the 6 elements we got above, we now have a complete basis for $su(3) =\mbox{span}\{(e_{Lpq} - e_{Lrs})\}$, where $p,q,r,s \in \{1,2,3,4,5,6\}$, and $e_{p}e_{q} = e_{r}e_{s}$.  Again, this generates the $SU(3)$ subgroup of $G_{2}$ that leaves $e_{7}$ invariant.  

The final element we get by taking the sum of the 3 elements above.  In particular, let 
$$
\mu = \frac{1}{6}(e_{L13}+e_{L26}+e_{L45}) = \frac{1}{6}(e_{L7} - e_{L7} + e_{L13}+e_{L26}+e_{L45}) = 
\frac{1}{6}(e_{L7} + 2e_{R7}).
$$
This element commutes with the elements of $su(3)$, and together they constitute a $u(3)$ subalgebra of $so(6)$, which is a subalgebra of our initial $so(7)$.  

In fact, however, we have four variations on this full reduction, each acting nontrivially on only one of the four subspaces of our spinor space \textbf{S}.  
$$
\begin{array}{l}
u(3)\rho_{L+}\rho_{R+} \; : \; {\bf{S}}_{++} \; \; (su(3) \mbox{ singlet}), \\
u(3)\rho_{L+}\rho_{R-} \; : \; {\bf{S}}_{+-} \; \; (su(3) \mbox{ triplet}), \\
u(3)\rho_{L-}\rho_{R+} \; : \; {\bf{S}}_{-+} \; \; (su(3) \mbox{ antitriplet}), \\
u(3)\rho_{L-}\rho_{R-} \; : \; {\bf{S}}_{--} \; \; (su(3) \mbox{ antisinglet}). \\
\end{array}
$$
And as to $\mu$, it has the following actions on the four subspaces:
$$
\begin{array}{l}
\mu{\bf{S}}_{++} = -\frac{i}{2}{\bf{S}}_{++}, \\
\mu{\bf{S}}_{+-} = +\frac{i}{6}{\bf{S}}_{+-}, \\
\mu{\bf{S}}_{-+} = -\frac{i}{6}{\bf{S}}_{-+}, \\
\mu{\bf{S}}_{--} = +\frac{i}{2}{\bf{S}}_{--}. \\
\end{array}
$$
Anyone familiar with the Standard Model of quarks and leptons will recognize this as the $u(1)$ hypercharge generator, or what can be interpreted as such.  
To this point it's just pure mathematics. \\

\newpage

\textbf{Complexified Quaternions as Spinors} \\

The algebras of left and right multiplicative actions of \textbf{C} on itself are identical, a result of its being both commutative and associative.  The algebras of left and right multiplicative actions of \textbf{O} on itself are also identical, a result of its being neither commutative nor associative.  \textbf{H}, however, in being associative, but not commutative, is in some ways more complicated than \textbf{C} or \textbf{O}.  While ${\bf{H}}_{L} \simeq {\bf{H}}_{R}$, these algebras are distinct, they commute with each other, and both are isomorphic to \textbf{H} itself.  The algebra of simultaneous left and right actions (${\bf{H}}_{A}$, 'A' for 'All') is isomorphic to ${\bf{R}}(4)$, the algebra of $4\times 4$ real matrices.

So \textbf{H} alone is the spinor space of ${\bf{H}}_{L} \simeq {\cal{CL}}(0,2)$, where there is an internal $SU(2)$ arising from ${\bf{H}}_{R}$, with respect to which the spinor space \textbf{H} is a doublet.  

But \textbf{H} alone can also be considered the spinor space of ${\bf{H}}_{A} \simeq {\cal{CL}}(2,2) \simeq {\cal{CL}}(3,1)$, in which case there is no algebra of internal actions.

Similar statements can be made about ${\bf{P}} = {\bf{C}}\otimes{\bf{H}}$.  ${\bf{P}}$ can be viewed as the spinor space of ${\bf{P}}_{L} \simeq {\cal{CL}}(3,0)$, which is isomorphic to the Pauli algebra, and where again ${\bf{H}}_{R}$ gives rise to an internal $SU(2)$; or ${\bf{P}}$ can be viewed as the spinor space of ${\bf{P}}_{A} \simeq {\cal{CL}}(0,5)$, which is also isomorphic to the Dirac algebra, \textbf{C}(4).  

The primary point I want to make here, one made ad nauseum in previous articles, is how the relation of \textbf{P} to that internal $SU(2)$ mentioned above differs from the relationship of \textbf{S} to the internal $SU(3)$ mentioned in the previous section.  To make this easier I'll look at ${\bf{P}}^{2}$ as the spinor space of ${\bf{P}}_{A}(2)$.  As usual I define a basis for ${\bf{R}}(2)$ with the following 4 matrices:
$$
\epsilon = 
\left[\begin{array}{rr}
1 & 0 \\ 0 & 1 \\
\end{array}\right], \,\,
\alpha = 
\left[\begin{array}{rr}
1 & 0 \\ 0 & -1 \\
\end{array}\right], \,\,
\beta = 
\left[\begin{array}{rr}
0 & 1 \\ 1 & 0 \\
\end{array}\right], \,\,
\gamma = 
\left[\begin{array}{rr}
0 & 1 \\ -1 & 0 \\
\end{array}\right].
$$
${\bf{P}}_{A}(2)$ is isomorphic to the complexification of ${\cal{CL}}(0,6)$, with a 1-vector basis 
$$
i q_{Lj} \gamma, \; \; q_{Rk} \beta, \;\; j,k \in \{1,2,3\}.
$$
The corresponding basis for the set of 2-vectors is 
$$
q_{Lj} \epsilon, \; \; q_{Rk} \epsilon, \;\; i q_{Lj}q_{Rk} \alpha, 
$$
which is the Lie algebra $spin(6) \simeq su(4)$.

I'm going to arbitrarily associate the 1-vectors $i q_{Lj} \gamma$ with Euclidean 3-space, 
and the 1-vectors $q_{Rk} \beta$ with extra dimensions.  The first thing I want to do is 
dimensionally reduce the 6-dimensional space to just Euclidean 3-space by using the matrix 
projectors
$$
E_{\pm} = \frac{1}{2} (\epsilon \pm i \gamma).
$$
Let's use the '+' projector, and note that 
$$
E_{+}i q_{Lj} \gamma E_{+} = i q_{Lj} \gamma E_{+}, 
$$
while 
$$
E_{+}q_{Rk} \beta E_{+} = E_{+}E_{-}q_{Rk} \beta = 0.
$$

At the 2-vector level, 
$$
\begin{array}{c}
E_{+}q_{Lj} \epsilon E_{+} = q_{Lj} \epsilon E_{+}, \\
E_{+}q_{Rk} \epsilon E_{+} = q_{Rk} \epsilon E_{+}, \\
E_{+}i q_{Lj}q_{Rk} \alpha E_{+} = 0, \\
\end{array}
$$
which is $spin(4) \simeq su(2)\times su(2)$, where one of the $su(2)$s is $spin(3)$ associated with 
the Euclidean space, which we'll denote $su_{L}(2)$ (or just $spin(3)$); and the other is internal 
(relative to 3-space), and we'll denote it $su_{R}(2)$.  

The spinor space, ${\bf{P}}^{2}$, is 16-dimensional, but $E_{+}{\bf{P}}^{2}$ is 8-dimensional, 
and relative to what we have left of the Clifford algebra constitutes an $SU(2)$ doublet of 
Pauli spinors.  As was the case for the spinor space ${\bf{C}}\otimes{\bf{O}}$, the ${\bf{C}}\otimes{\bf{H}}$ 
parts of this new spinor space have an algebraic structure, and again admit a nontrivial resolution 
of the identity into a pair of orthogonal projection operators.  Our choice here is the following:
$$
\lambda_{\pm} = \frac{1}{2}(1 \pm iq_{3})
$$
(where $q_{k}, k = 1,2,3$, are the three imaginary quaternionic units, which I hope by now is obvious).  We 
can use these to further decompose $E_{+}{\bf{P}}^{2}$:
$$
\begin{array}{ll}
\lambda_{+}E_{+}{\bf{P}}^{2}\lambda_{+} & \lambda_{+}E_{+}{\bf{P}}^{2}\lambda_{-} \\
\lambda_{-}E_{+}{\bf{P}}^{2}\lambda_{+} & \lambda_{-}E_{+}{\bf{P}}^{2}\lambda_{-} \\
\end{array}
$$
These 4 spinor components are just single complex numbers (along with the projectors).  However (and this is a huge contrast to the 
${\bf{C}}\otimes{\bf{O}}$ case), $\lambda_{\pm}$ acting as a projection on the spinor from the left only 
effects that part of the spinor associated with $spin(3)$, while $\lambda_{\pm}$ projecting from the right 
pick out the two halves of the $su_{R}(2)$ spinor $E_{+}{\bf{P}}^{2}$.  

My point is to contrast the internal $su(3)$ derived above with this new internal $su(2)$ associated 
with ${\bf{H}}_{R}$, so the next step is to pull the action 
$$
E_{+}{\bf{P}}^{2}\lambda_{\pm} = \lambda_{R\pm}[E_{+}{\bf{P}}^{2}]
$$
back to the Clifford algebra, which looks like 
$$
\lambda_{R\pm}E_{+}{\cal{CL}}(0,6)E_{+}\lambda_{R\pm}
$$
(my point is made better by using the Clifford algebra already reduced by $E_{+}$).  Our remaining 
three 1-vectors are linear in $q_{Lj}$, which commute with any element of ${\bf{H}}_{R}$, so these 
elements remain unchanged by the $\lambda_{R\pm}$ reduction above.  The same holds true at the 
2-vector level for $su_{L}(2) \simeq spin(3)$.  On the other hand, 
$$
\begin{array}{l}
\lambda_{R\pm}q_{Rk} \epsilon\lambda_{R\pm} = 0, \;\; k = 1,2, \\
\lambda_{R\pm}q_{Rk} \epsilon\lambda_{R\pm} = \mp i\lambda_{R\pm}\epsilon, \;\; k = 3. \\
\end{array}
$$
That is, $\lambda_{R\pm}$ reduces the internal $su(2)$ to $u(1)$; it breaks the symmetry.  

With ${\bf{C}}\otimes{\bf{O}}$ the projectors left us with $u(3)$, an exact part of the standard 
symmetry (Lie algebra) $u(1)\times su(2)\times su(3)$.  This $su(2)$ is not exact, but has a 
$u(1)$ subalgebra which is, and this is what we're left with in the  ${\bf{C}}\otimes{\bf{H}}$ case.  
This is not to say that we have yet a sufficiently developed picture to make the claim that we now 
have the standard model, only that the symmetries of that model are inherently a part of this picture, 
as is a natural mechanism for yielding an exact $u(3)$ and broken $su(2)$.  It's right there in the 
mathematics.  See it?  Right there ... no, right over there.  It's purple ... look for something 
purple.  Just to conclude this section: the projectors $\rho_{\pm}$ projected from the spinor space 
${\bf{C}}\times{\bf{O}}$ entire multiplets (${\bf{1}}\oplus{\bf{3}}\oplus\overline{{\bf{1}}}\oplus 
\overline{{\bf{3}}}$), while the $\lambda_{R\pm}$ projected from an extant $su(2)$ doublet its 
individual components. \\

\textbf{All Three} \\

Define
$$
{\bf{T}} \equiv {\bf{C}}\otimes{\bf{H}}\otimes{\bf{O}}.
$$
This is the spinor space of the Clifford algebras
$$
{\bf{T}}_{L} \simeq {\cal{CL}}(0,9)
$$
and
$$
{\bf{T}}_{A} \simeq {\cal{CL}}(11,0),
$$
where ${\bf{T}}_{L}$ uses only ${\bf{H}}_{L}$, and ${\bf{T}}_{A}$ uses ${\bf{H}}_{A}$, which includes 
both ${\bf{H}}_{L}$, ${\bf{H}}_{R}$, and their combined actions.  As I went to the trouble of writing 
a book largely about \textbf{T} ([1]), and many articles ([2][3]), I will here summarize the results.  

\textbf{T} inherits noncommutativity from \textbf{H} and \textbf{O}, and nonassociativity from \textbf{O}.  
From the combination of \textbf{H} and \textbf{O} it also loses alternativity.  That is, there are elements 
x and y such that $x(xy) \ne (x^2)y$.  Worse, there are idempotents $x$ that do not alternate.  That is, 
there exist $y$ such that 
$$
x(xy) \ne (x^2)y = xy.
$$
Such idempotents obey the letter of the law (definition) of idempotent, but not its spirit (intention), 
which really is an element $x$ of an algebra {\bf{A}} satisfying, to begin with, 
$$
x(xy) = xy
$$
for all $y \in \bf{A}$.  That is, x is a projection operator.  As to ${\bf{T}}$, there a several ways 
to resolve its identity into four orthogonal \textit{idempotents} ($\Delta_{m}, \; m= 0,1,2,3$), but only one way 
(I believe), up to automorphism, satisfying for all y in \textbf{T},
$$
\begin{array}{l}
\Delta_{0} + \Delta_{1} + \Delta_{2} + \Delta_{3} = 1, \\
\Delta_{m}(\Delta_{n}y) = \delta_{mn}\Delta_{m}y, \\
(y\Delta_{m})\Delta_{n} = \delta_{mn}y\Delta_{m}, \\
\Delta_{m}(y\Delta_{n}) = (\Delta_{m}y)\Delta_{n}, \\
\end{array}
$$
these properties ensuring the $\Delta_{m}$ can be treated as a set of orthogonal projection operators 
with which we can consistently decompose the spinor space \textbf{T}.

Drawn from the book [1], we make the assignments:
$$
\begin{array}{l}
\Delta_{0} = \frac{1}{4}(1+i\vec{x})(1+ie_{7}) = (\frac{1}{2}(1+i\vec{x}))(\frac{1}{2}(1+ie_{7})) = \lambda_{0}\rho_{+}, \\
\Delta_{1} = \frac{1}{4}(1-i\vec{x})(1+ie_{7}) = (\frac{1}{2}(1-i\vec{x}))(\frac{1}{2}(1+ie_{7})) = \lambda_{1}\rho_{+}, \\
\Delta_{2} = \frac{1}{4}(1+i\vec{y})(1-ie_{7}) = (\frac{1}{2}(1+i\vec{y}))(\frac{1}{2}(1-ie_{7})) = \lambda_{2}\rho_{-}, \\
\Delta_{3} = \frac{1}{4}(1-i\vec{y})(1-ie_{7}) = (\frac{1}{2}(1-i\vec{y}))(\frac{1}{2}(1-ie_{7})) = \lambda_{3}\rho_{-}, \\
\end{array}
$$
where $\vec{x}$ and $\vec{y}$ are linear in $q_{k}, k=1,2,3$ and independent (see [1]).  The $\lambda_{m}$ are similar 
to the $\lambda_{\pm}$ defined above, and similar things can be said about them.  For example, they provide a mechanism for 
breaking $SU(2)$, whereas nothing in the $\Delta_{m}$ can break $SU(3)$.  Moreover, the $\Delta_{m}$ themselves are 
elements of \textbf{T}, and they are invariant with respect to $SU(3)$, but not $SU(2)$.  In [1] this helped provide an 
explanation for the nonchirality of $SU(3)$ and the chirality of $SU(2)$.

One other thing that occurred to me in Denver.  As I see it, theories that require some sort of 
representation or other structure to explain the organization of elementary particles, often do 
an okay job of this.  For example, it was noted long ago, that $SU(3)$ did great things and the 
\textbf{3} multiplet suggested the existence of quarks, which are fermions, therefore presumably individually 
describable to some extent by Dirac spinors.  Well, the $SU(3)$ triplet is not inherently a triplet 
of Dirac (or Pauli) spinors.  You have to stick that structure on, like a fiber.  In our case we 
start with a spinor - a more fundamental structure than the inherently bosonic spacetime - and our 
particular spinor spaces have associated algebraic structures that give rise to associated Lie groups 
and a natural multiplet structure within the spinor.  \textbf{T}, for example, is a Pauli spinor 
doublet for a 1,9-spacetime in exactly the same way \textbf{P} is a Pauli spinor doublet for 
1,3-spacetime.  \textbf{P} can be viewed as a doublet of ordinary Pauli spinors in ${\bf{C}}^{2}$.  
And how about the ordinary ${\bf{C}}^{2}$ Pauli spinors in \textbf{T}?  There is an entire lepto-quark 
family and anti-family of them.  \textbf{T} is an example of an all-inclusive hyperspinor, one that 
carries within its mathematical self the keys one needs to view it within the context of 1,3-spacetime.  \\

\textbf{Final Repetitive Thoughts} \\

Let $\Psi$ be a ${\bf{T}}^{2}$ spinor with a dependence on 1,9-spacetime, a la [1].  As shown in that 
reference, we can easily pick out of this [family]+[antifamily] hyperspinor parts that are associated 
with recognizable lepton fields.  For example, 
$$
\rho_{+}\Psi\rho_{+}\lambda_{0} = \rho_{+}\Psi\Delta_{0} \longrightarrow \mbox{Dirac spinor for a neutrino field}, 
$$
where we know this is a neutrino because it has all the charges a neutrino field ought to have with respect 
to $U(3)$ and $SU(2)$.  We can make a simple Dirac-like Lagrangian for these fields (see [1] - it's more 
complicated than this):
$$
{\cal{L}} = \overline{\Psi} \not{\partial} \Psi,
$$
where the parametrization of the fields in this case is over 1,9-spacetime.  This can be gauged, and the 
final result has a really nice property: all allowable interactions can be read from the real part of 
$\cal{L}$.  For example, let $\nu$ be the neutrino part of $\Psi$, $u_{g}$ be the green up-quark part 
of $\Psi$ (green, of course, being shorthand for one set of the up-quarks $SU(3)$ charges, or colors), 
and $W^{+}$ be an intermediate vector boson arising from the gauging of $SU(2)$.  Then this term exists 
in $\cal{L}$,  
$$
\overline{\nu} W^{+} u_{g},
$$
but it is not in the real part of $\cal{L}$, therefore it is not an allowable interaction.  No QFT with 
Feynman diagrams is needed at this point to determine all the interactions that can happen.  The fields, 
their charges, the consequent particle identifications, and the particle interactions, are all inherent 
in the mathematics of the division algebras within \textbf{T} (see [1][2][3][4] for much more elaborate 
discussions of these matters).

And now something you won't find in the references.  There are 3 parallelizable spheres: $S^{1}$, $S^{3}$, 
and $S^{7}$, and no others.  The parallelizability can give rise to a product structure on each of these 
spheres, essentially deriving the subalgebras of \textbf{C}, \textbf{H} and \textbf{O}, consisting of their 
unit elements.  Then, with these product structures, 
$$
{\bf{T}} = {\bf{R}}\otimes S^{1} \otimes S^{3} \otimes S^{7}.
$$
I mention this because one might wonder why we need to include \textbf{C} and \textbf{H}, since they are 
already contained in \textbf{O} as subalgebras.  This is a disease almost all mathematical physicists have, 
assigning fundamental significance to human defined structures.  \textbf{C}, \textbf{H} and \textbf{O}, are 
much more than a series of \textit{algebras}, they are a series of mathematical universes, each with very 
different properties, and for each of these mathematical universes the corresponding division algebra is 
just a sign post that says: Enter Here.  Without the inclusion of each of these separate mathematical 
universes you do not (can not?) get the 1-to-1 correspondence of the mathematics to the particles, fields, 
and interactions that we observe in our physical universe.  And if your model does an excellent job of 
doing this, then either the universe is playfully perverse, or your model has \textbf{T} hidden in it, 
perhaps in some non-obvious way. \\

\textbf{References}: \,

[1] G.M. Dixon, \textit{Division Algebras: Octonions, Quaternions, Complex Numbers, and the Algebraic Design of Physics}, (Kluwer (Now Springer), 1994). 

[2] G.M. Dixon, http://www.7stones.com 

[3] G.M. Dixon, J. Math. Phys. 45, 3878 (2004) 

[4] G.M. Dixon, http://www.7stones.com/Homepage/10Dnew.pdf  \\
.\hspace{.36in}G.M. Dixon, http://www.7stones.com/Homepage/10parity.pdf  \\
.\hspace{.36in}G.M. Dixon, http://www.7stones.com/Homepage/14mix.pdf \\
.\hspace{.36in}G.M. Dixon, http://www.7stones.com/Homepage/6x6.pdf \\
.\hspace{.36in}G.M. Dixon, http://www.7stones.com/Homepage/123cho.pdf 

[5] John Baez, The octonions, Bull. Amer. Math. Soc. 39 (2002), 145-205 \\

\end{document}